\documentclass[preprint2]{aastex6}

\mathchardef\mhyphen="2D 

\begin{document}

\title{A non-LTE study of silicon abundances in giant stars from the \ion{Si}{1} infrared lines in the zJ-band\altaffilmark{1}}

\author{Kefeng Tan and Jianrong Shi}\affil{Key Laboratory of Optical Astronomy, National Astronomical Observatories,
Chinese Academy of Sciences, Beijing 100012, China}\email{tan@nao.cas.cn}

\author{Masahide Takada-Hidai}\affil{Liberal Arts Education Center, Tokai University, 4-1-1 Kitakaname, Hiratsuka, Kanagawa 259-1292, Japan}

\author{Yoichi Takeda}\affil{National Astronomical Observatory of Japan, 2-21-1 Osawa, Mitaka, Tokyo 181-8588, Japan}

\and

\author{Gang Zhao}\affil{Key Laboratory of Optical Astronomy, National Astronomical Observatories,
Chinese Academy of Sciences, Beijing 100012, China}

\altaffiltext{1}{Based on data collected at Subaru Telescope, which is operated by the
National Astronomical Observatory of Japan; based on observations made with ESO telescopes at
the La Silla Paranal Observatory under programme IDs 266.D-5655(A) and 084.D-0912(A); based on
observations carried out at the National Astronomical Observatories (Xinglong, China).}

\begin{abstract}
  We investigate the feasibility of the \ion{Si}{1} infrared (IR) lines as Si abundance indicators
  for giant stars. We find that Si abundances obtained from the \ion{Si}{1} IR lines
  based on the local thermodynamic equilibrium (LTE) analysis show large line-to-line scatter
  (mean value of 0.13\,dex), and are higher than those from the optical lines. However, when
  the non-LTE effects are taken into account, the line-to-line scatter reduces significantly
  (mean value of 0.06\,dex), and the Si abundances are consistent with those from the optical
  lines. The typical average non-LTE correction of [Si/Fe] for our sample stars is about $-0.35$\,dex.
  Our results demonstrate that the \ion{Si}{1} IR lines could be reliable abundance indicators
  provided that the non-LTE effects are properly taken into account.
\end{abstract}


\section{Introduction}\label{intro}

Chemical composition of stellar photospheres is a very important tool to investigate the
origins of elements as well as the chemical evolution of the galaxies. During the past
decades, stellar abundances\footnote{We follow the classical notation
$\mathrm{[A/B]}=\log(N_{\mathrm{A}}/N_{\mathrm{B}})_{\bigstar}-\log(N_{\mathrm{A}}/N_{\mathrm{B}})_{\odot}$
in this work.} are mainly determined from optical spectra due to both historical and
technical reasons. Though visible light can easily penetrate the atmosphere of the earth,
they are subject to the obscuring of interstellar dust and gas. This makes it difficult
to utilize the optical spectra to investigate the chemical abundances of stars suffering
heavy interstellar extinction, such as stars in the inner Galactic disk and in the Galactic
bulge. In this regard, infrared (IR) spectroscopy is more promising as they are much less
affected by the interstellar extinction. The ongoing Apache Point Observatory Galactic
Evolution Experiment (APOGEE; \citealt{apogee}) is such an attempt to use high-resolution
and high signal-to-noise (S/N) ratio IR spectroscopy to penetrate the dust that obscures
significant fractions of the disk and bulge of the Galaxy. Furthermore, IR spectroscopy
could even be used to investigate the chemical abundances of stars beyond our Galaxy.
Owning to their large luminosities and peak flux in the IR, red supergiant (RSG) stars
are ideal tracers of the chemical abundances of the external galaxies out to large
distances (\citealt{patrick15}). This has been verified by chemical abundance analysis
of RSG stars in the Magellanic Clouds (\citealt{davies15}), in NGC\,6822
(\citealt{patrick15}), and in the Sculptor Galaxy (\citealt{gazak15}) using the
medium-resolution ($R\sim3000$--8000) spectra in the $J$-band obtained by X-shooter
(\citealt{xshooter}) and $K$-band Multi-Object Spectrograph
(KMOS; \citealt{kmos}) mounted on the Very Large Telescope. \citet{evans11} showed by
simulation that, with future instruments, quantitative IR spectroscopy could even be
performed for RSG stars to tens of Mpc. Due to the aforementioned advantages, IR
spectroscopy will play a more important role in chemical abundance analysis with existing
and forthcoming IR instruments (such as Keck/MOSFIRE, VLT/KMOS, TMT/IRMS, E-ELT/EAGLE, etc.).

In addition to the advantages mentioned above, sometimes IR lines could be better abundance
indicators than optical lines. This is the case for silicon (Si) in very metal-poor stars.
Si is attributed as an $\alpha$-element, which is made during oxygen and neon
burning in massive stars, and later ejected to the interstellar medium by Type II supernovae
(SNe) according to \citet{woosley95}. SNe~Ia may also produce some Si as suggested by
\citet{tsujimoto95}. Therefore, Si abundances in metal-poor stars could be used to test the
SNe and Galactic chemical evolution models.
Unfortunately, the optical \ion{Si}{1} lines are very weak in very metal-poor stars, so the
two strong lines at 3905 and 4102\,{\AA} in the near-ultraviolet (NUV) are usually employed
to derive Si abundances. However, both of these two lines have defects in abundance
determinations. The \ion{Si}{1} 3905\,{\AA} line is blended by a CH line. Though this CH
feature may be weak in dwarf stars with relatively high temperature (\citealt{cohen04}), it
could be very strong in cool giant stars (\citealt{cayrel04}). The \ion{Si}{1} 4102\,{\AA}
line falls in the wing of the H$\delta$ line, which makes it not easy to derive accurate Si
abundance.
In such a situation, the \ion{Si}{1} IR lines could be a better alternative to derive Si
abundances. There are tens of \ion{Si}{1} IR lines that are much stronger than the optical
lines and they suffer much less from the problem of blending or continuum normalization
compared to the 3905/4102\,{\AA} lines. For example, \citet{jonsson11} determined Si abundances for ten
metal-poor giant stars using three \ion{Si}{1} IR lines at 10371, 10844, and 10883\,{\AA}.
However, their analysis was under the assumption of local thermodynamic equilibrium (LTE)
as there was no calculation for the non-LTE effects of the \ion{Si}{1} IR lines at that time.
Later, \citet{shi12} investigated the non-LTE effects of the \ion{Si}{1} IR lines in nearby
stars (most of which were dwarf stars), and found that the non-LTE effects are important
even for metal-rich stars ($>0.1$\,dex). \citet{bergemann13} presented
theoretical calculations for the non-LTE effects of four \ion{Si}{1} lines in the $J$-band
for RSG stars. Their results show that the non-LTE abundance correction varies smoothly
between $-0.4$ and $-0.1$\,dex for stars with effective temperature between 3400 and 4400\,K.

Considering that \citet{shi12} concentrated mainly on main-sequence stars, while
\citet{bergemann13} focused on the theoretical non-LTE effects, we decided to make a
practical investigation on the non-LTE effects of 16 \ion{Si}{1} IR lines in giant stars
based on observational spectra. In particular, the atomic data and model atom of
\citet{shi12} were calibrated by requiring that consistent Si abundances could be
obtained from different \ion{Si}{1} IR lines as well as from the optical lines for the sun,
and it is necessary to check whether these still hold true for giant stars.
In the next section, we briefly describe the observational data used in this work.
Section~\ref{method_test} presents our method of non-LTE calculations and the test of its
validity for giants. In Section~\ref{appl}, we apply our method to a sample of metal-poor
giant stars and compare our results with the theoretical chemical evolution models of Si.
In the last section we briefly summarize our results and conclusions.

\section{Observational data}\label{obs}

\floattable
\begin{deluxetable}{lccCccrcccc}
\tablewidth{0pt}
\tablecolumns{10}
\tablecaption{Sample stars and their stellar parameters. From left to right: star ID, effective temperature, surface gravity, metallicity,
microturbulent velocity, reference for stellar parameters, adopted uncertainty for effective temperature, surface gravity, metallicity,
and microturbulent velocity, references or note about the calculation of the uncertainty of stellar parameters. For stars with multiple
independent measurements of stellar parameters in the references, the standard deviations between different studies were adopted as the
uncertainties; for the other stars please see the note in the end of the table.\label{table:sample}}
\tablehead{\colhead{Star} & \colhead{$T_{\mathrm{eff}}$} & \colhead{$\log g$} & \colhead{[Fe/H]} & \colhead{$\xi$} & \colhead{Ref.}
 & \colhead{$\sigma_t$} & \colhead{$\sigma_g$} & \colhead{$\sigma_m$} & \colhead{$\sigma_v$} & \colhead{Ref./Note}\\
 & \colhead{(K)} & \colhead{(cgs)} & \colhead{(dex)} & \colhead{(km\,s$^{-1}$)} & & \colhead{(K)} & \colhead{(cgs)} & \colhead{(dex)} & \colhead{(km\,s$^{-1}$)} & }
\startdata
Arcturus            & 4281 & 1.72 & -0.55 & 1.5 & a &  20 & 0.08 & 0.07 & 0.25 & a\\
HD\,83240           & 4682 & 2.45 & -0.02 & 1.3 & b &  84 & 0.27 & 0.01 & 0.03 & b, l\\
\tableline
BD\,$+23\degr 3130$ & 5000 & 2.20 & -2.60 & 1.4 & c &  71 & 0.22 & 0.14 & 0.28 & $\star$\\
BD\,$-16\degr 251$  & 4825 & 1.50 & -2.91 & 1.8 & d &  71 & 0.01 & 0.07 & 0.28 & d, i\\
BD\,$-18\degr 5550$ & 4750 & 1.40 & -3.06 & 1.8 & d & 124 & 0.00 & 0.04 & 0.00 & d, f\\
HD\,6268            & 4735 & 1.61 & -2.30 & 2.1 & e &  25 & 0.01 & 0.45 & 0.35 & e, f\\
HD\,13979           & 5075 & 1.90 & -2.26 & 1.3 & f &  71 & 0.22 & 0.14 & 0.28 & $\star$\\
HD\,108317          & 5310 & 2.77 & -2.35 & 1.9 & g & 100 & 0.15 & 0.09 & 0.46 & c, f, g, h, i\\
HD\,115444          & 4721 & 1.74 & -2.71 & 2.0 & h &  32 & 0.52 & 0.12 & 0.32 & e, h, i\\
HD\,121135          & 4934 & 1.91 & -1.37 & 1.6 & h &   6 & 0.29 & 0.03 & 0.28 & f, h\\
HD\,126587          & 4700 & 1.05 & -3.16 & 1.7 & i &  18 & 0.88 & 0.25 & 0.11 & c, i\\
HD\,166161          & 5350 & 2.56 & -1.22 & 2.3 & h & 100 & 0.22 & 0.14 & 0.18 & f, h, i\\
HD\,186478          & 4730 & 1.50 & -2.42 & 1.8 & i &  76 & 0.08 & 0.18 & 0.10 & d, f, h, i\\
HD\,195636          & 5370 & 2.40 & -2.77 & 1.5 & j &  71 & 0.22 & 0.14 & 0.28 & $\star$\\
HD\,204543          & 4672 & 1.49 & -1.72 & 2.0 & h &  16 & 0.47 & 0.21 & 0.00 & f, h, i\\
HD\,216143          & 4525 & 1.77 & -1.92 & 1.9 & e &   8 & 0.51 & 0.06 & 0.48 & c, e, f, g\\
HD\,221170          & 4560 & 1.37 & -2.00 & 1.6 & e &  74 & 0.22 & 0.14 & 0.52 & c, e, f, g, h\\
HE\,1523$-$0901     & 4630 & 1.00 & -2.95 & 2.6 & k &  71 & 0.22 & 0.14 & 0.28 & $\star$\\
\enddata
\tablerefs{a: \citet{takeda09}; b: \citet{mishenina06}; c: \citet{fulbright00}; d: \citet{cayrel04}; e: \citet{saito09};
f: \citet{burris00}; g: \citet{takada05}; h: \citet{simmerer04}; i: \citet{hansen11}; j: \citet{carney03}; k: \citet{frebel07};
l: \citet{dasilva11}.}
\tablecomments{$\star$ Only single independent measurement of stellar parameters was available in the references; the median value
of the errors of the 12 stars with multiple independent measurements of stellar parameters was adopted as the uncertainties.}
\end{deluxetable}

The sample analyzed in this work comprises 16 metal-poor giant program stars as shown in
Table~\ref{table:sample}. These stars were originally observed by \citet{takeda11,takeda12}
in 2009 and 2011 for the purpose of determining sulfur (S) abundances using the \ion{S}{1}
IR triplet lines. The spectra were obtained using the IR Camera and Spectrograph
(IRCS; \citealt{ircs}) in combination with the 188-element curvature-based adaptive optics
system (AO188) mounted on the Subaru Telescope. With a resolution of about 20,000, the
spectra cover a wavelength range of 1.01--1.19\,$\mu$m, including several \ion{Si}{1}
lines that are not (severely) blended. For most of the stars, the S/N ratios of
the spectra are higher than 100. The spectra were reduced following the standard procedure
using the ``echelle'' package of IRAF\footnote{IRAF is distributed by the National Optical
Astronomy Observatories, which are operated by the Association of Universities for Research
in Astronomy, Inc., under cooperative agreement with the National Science Foundation.}. For
more details about the observation and data reduction, please refer to \citet{takeda11,takeda12}.

As mentioned in the introduction, one of our aims is to check whether we could obtain
consistent Si abundances from the IR and the optical lines in giant stars. Unfortunately,
the optical \ion{Si}{1} lines are too weak to give very accurate abundances for our sample
stars. So we included another two giant stars (Arcturus and HD\,83240) as benchmark stars.
The optical \ion{Si}{1} lines in these two stars are stronger compared to the 16 sample
stars, which permits us to check whether abundances from the optical and the IR lines are
in reasonable agreement. Besides, these two stars have archival IR spectra of very high
quality, based on which we could obtain very accurate Si abundances to investigate whether
different IR lines produce consistent results. For Arcturus, the optical spectra was
adopted from the Visible and Near IR Atlas of the Arcturus Spectrum 3727--9300\,{\AA} by
\citet{hinkle00}, and the IR spectra was adopted from the IR Atlas of the Arcturus Spectrum,
0.9--5.3 microns by \citet{hinkle95}. The typical spectral resolution of the optical and IR
spectra of Arcturus is 150,000 and 100,000, respectively. For HD\,83240, the optical spectra
was adopted from the UVES-POP library (\citealt{uvespop}), and the IR spectra was adopted
from the CRIRES-POP library (\citealt{crirespop}). The typical spectral resolution of the
optical and the IR spectra of HD\,83240 is 80,000 and 96,000, respectively.

\section{Method of non-LTE calculations and validity test}\label{method_test}

\subsection{Method of non-LTE calculations}\label{method}

\floattable
\begin{deluxetable}{rr@{$-$}lcrc}
\tablewidth{0pt}
\tablecolumns{6}
\tablecaption{Atomic data of the \ion{Si}{1} lines (adopted from \citealt{shi08}) used for abundance determination. From left to right:
wavelength, lower and upper level of the transition, excitation potential (EP) of the lower level, oscillator strength, van der Waals
damping constant.\label{table:line}}
\tablehead{\colhead{Line ({\AA})} & \multicolumn{2}{c}{Transition} & \colhead{EP (eV)} & \colhead{$\log gf$} & \colhead{$\log C_6$}}
\startdata
 5690.43 & $\mathrm{4s\,^3P^o_1}$   & $\mathrm{5p\,^3P_1}$   & 4.707 & $-1.73$ & $-30.294$\\
 5701.11 & $\mathrm{4s\,^3P^o_1}$   & $\mathrm{5p\,^3P_0}$   & 4.707 & $-1.95$ & $-30.294$\\
 6142.49 & $\mathrm{3p^3\,^3D^o_3}$ & $\mathrm{5f\,^3D_3}$   & 5.619 & $-1.47$ & $-29.869$\\
 6145.02 & $\mathrm{3p^3\,^3D^o_2}$ & $\mathrm{5f\,^3G_3}$   & 5.616 & $-1.38$ & $-29.869$\\
\tableline
10288.90 & $\mathrm{4s\,^3P^o_0}$   & $\mathrm{4p\,^3S_1}$   & 4.920 & $-1.65$ & $-30.661$\\
10371.30 & $\mathrm{4s\,^3P^o_1}$   & $\mathrm{4p\,^3S_1}$   & 4.707 & $-0.85$ & $-30.659$\\
10585.17 & $\mathrm{4s\,^3P^o_2}$   & $\mathrm{4p\,^3S_1}$   & 4.954 & $-0.14$ & $-30.659$\\
10603.45 & $\mathrm{4s\,^3P^o_1}$   & $\mathrm{4p\,^3P_2}$   & 4.707 & $-0.34$ & $-30.677$\\
10627.66 & $\mathrm{4p\,^1P_1}$     & $\mathrm{4d\,^3P^o_2}$ & 5.863 & $-0.39$ & $-30.692$\\
10661.00 & $\mathrm{4s\,^3P^o_0}$   & $\mathrm{4p\,^3P_1}$   & 4.920 & $-0.28$ & $-30.687$\\
10689.73 & $\mathrm{4p\,^3D_1}$     & $\mathrm{4d\,^3F^o_2}$ & 5.954 & $-0.08$ & $-29.964$\\
10694.27 & $\mathrm{4p\,^3D_2}$     & $\mathrm{4d\,^3F^o_3}$ & 5.964 & $ 0.06$ & $-29.944$\\
10727.43 & $\mathrm{4p\,^3D_3}$     & $\mathrm{4d\,^3F^o_4}$ & 5.984 & $ 0.25$ & $-29.907$\\
10749.40 & $\mathrm{4s\,^3P^o_1}$   & $\mathrm{4p\,^3P_1}$   & 4.707 & $-0.20$ & $-30.689$\\
10784.57 & $\mathrm{4p\,^3D_2}$     & $\mathrm{4d\,^3F^o_2}$ & 5.964 & $-0.69$ & $-29.965$\\
10786.88 & $\mathrm{4s\,^3P^o_1}$   & $\mathrm{4p\,^3P_0}$   & 4.707 & $-0.34$ & $-30.691$\\
10827.10 & $\mathrm{4s\,^3P^o_2}$   & $\mathrm{4p\,^3P_2}$   & 4.954 & $ 0.21$ & $-30.677$\\
10843.87 & $\mathrm{4p\,^1P_1}$     & $\mathrm{4d\,^1D^o_2}$ & 5.863 & $-0.08$ & $-30.145$\\
10882.83 & $\mathrm{4p\,^3D_3}$     & $\mathrm{4d\,^3F^o_3}$ & 5.984 & $-0.66$ & $-29.945$\\
10979.34 & $\mathrm{4s\,^3P^o_2}$   & $\mathrm{4p\,^3P_1}$   & 4.954 & $-0.55$ & $-30.688$\\
\enddata
\end{deluxetable}

In this work, we used the same method of non-LTE calculations as \citet{shi08,shi09,shi11,shi12}.
Here, we only give a brief description of the method. The key of non-LTE calculations is
the atomic model. The Si atomic model adopted here includes 132 terms of \ion{Si}{1}, 41
terms of \ion{Si}{2}, and the ground state of \ion{Si}{3}. In addition to the radiative
bound--bound and bound--free transitions, excitation and ionization induced by inelastic
collisions with electrons and hydrogen atoms were also taken into account. The atomic data
of the spectral lines used for Si abundance determination are given in Table~\ref{table:line}.
The data were adopted from \citet{shi08}, where the van der Waals damping constants were
computed according to the interpolation tables of \citet{anstee91,anstee95}, and the
oscillator strengths were derived by requiring the solar Si abundance
$\log(N_{\mathrm{Si}}/N_{\mathrm{H}})+12$ to be 7.5. As for the stellar model atmosphere,
we used the revised version of the opacity sampling model MAFAGS-OS (\citealt{grupp04,grupp09}),
which is under the one-dimensional (1D) and LTE assumption. The coupled radiative transfer
and statistical equilibrium equations were solved with a revised version of the DETAIL
program (\citealt{detail}). Chemical abundances were derived by spectrum synthesis
using the IDL/FORTRAN-based software package Spectrum Investigation Utility (SIU)
developed by Dr.~J. Reetz. SIU is an advanced software package which could perform
spectrum displaying, continuum normalization, radial velocity correlation/correction,
spectrum synthesis (in LTE or non-LTE based on the 1D-LTE model atmosphere MAFAGS-OS
mentioned above), etc., interactively. All the input parameters, such as abundance for
each element, macroturbulence, rotation, and instrument broadening, could be adjusted
interactively, and the synthetic spectrum could be displayed on screen in real time. In
most cases, the external broadening of line profiles caused by macroturbulence, rotation,
and instrument could be approximated using a simple Gaussian function.

\subsection{Validity test for giants with benchmark stars}\label{test}

The above method of non-LTE calculations has been proved to be applicable for dwarf stars
by \citet{shi12}. In this section we will test whether it is also valid for giant stars.

\subsubsection{Stellar parameters and uncertainties}

For Arcturus, we adopted the stellar parameters ($T_{\mathrm{eff}}=4281$\,K, $\log g=1.72$,
$\mathrm{[Fe/H]}=-0.55$, $\xi=1.5$\,km\,s$^{-1}$) from \citet{takeda11}. They were
determined by \citet{takeda09} in a full spectroscopic way, i.e., $T_{\mathrm{eff}}$ from
excitation equilibrium of \ion{Fe}{1}, $\log g$ from ionization equilibrium between
\ion{Fe}{1} and \ion{Fe}{2}, [Fe/H] from \ion{Fe}{1}, and $\xi$ from abundance-independence
of equivalent width (EW). We noticed that \citet{ramirez11} also investigated the stellar
parameters and abundances of Arcturus using high-quality data and methods that minimize
model uncertainties. They determined the effective temperature using model atmosphere fits
to the observed spectral energy distribution from the blue to the mid-IR, and the surface
gravity using the trigonometric parallax. Their stellar parameters
($T_{\mathrm{eff}}=4286\pm30$\,K, $\log g=1.66\pm0.05$, $\mathrm{[Fe/H]}=-0.52\pm0.04$,
$\xi=1.74$\,km\,s$^{-1}$) are in excellent agreement with those adopted in this work.
Recently, \citet{heiter15} presented fundamental determinations of $T_{\mathrm{eff}}$ and
$\log g$ for 34 \emph{Gaia} FGK benchmark stars. Their determinations were based on the
angular diameters, bolometric fluxes, distances, and masses, which are independent of
spectroscopy and atmospheric models. They gave a $T_{\mathrm{eff}}$ of $4286\pm35$\,K and
a $\log g$ of $1.64\pm0.09$ for Arcturus. And the [Fe/H] of this star was determined to be
$-0.52\pm0.08$ (after correction for non-LTE effects) in another work (\citealt{jofre14})
in the \emph{Gaia} FGK benchmark stars series. These values agree well with the two sets
of stellar parameters mentioned above, and the offsets between different studies are well
within the typical uncertainties of stellar parameters ($\Delta T_{\mathrm{eff}}=20$\,K,
$\Delta\log g=0.08$, $\Delta\mathrm{[Fe/H]}=0.07$, and $\Delta\xi=0.25$\,km\,s$^{-1}$)
given by \citet{takeda09}.
Actually, our calculations show that such differences in stellar parameters had a very
small effect on the Si abundance ($\Delta\mathrm{[Si/Fe]}\sim0.04$\,dex).

For HD\,83240, we adopted the stellar parameters ($T_{\mathrm{eff}}=4682$\,K, $\log g=2.45$,
$\mathrm{[Fe/H]}=-0.02$, $\xi=1.3$\,km\,s$^{-1}$) from \citet{mishenina06}. The effective
temperature was determined with very high accuracy ($\sigma=10$--15\,K) using the line
depth ratios of several iron-peak elements (such as Si, Ti, V, Cr, Fe, and Ni). The
surface gravity was derived using two methods, i.e., iron ionization equilibrium and wing
fitting of the \ion{Ca}{1} 6162\,{\AA} line, and both methods give the same result. The
metallicity was determined from \ion{Fe}{1} lines. We note that according to the
investigation of \citet{lind12}, non-LTE effects of \ion{Fe}{1} are negligible for cool
($T_{\mathrm{eff}}<5000$\,K) stars with solar metallicity ($\mathrm{[Fe/H]}\sim0$), such
as HD\,83240. The microturbulent velocity was calculated so that iron abundances do not
depend on the EWs. \citet{dasilva11} gave different stellar parameters for HD\,83240
($T_{\mathrm{eff}}=4801\pm89$\,K, $\log g=2.83\pm0.23$, $\mathrm{[Fe/H]}=-0.03\pm0.08$,
$\xi=1.26\pm0.09$\,km\,s$^{-1}$) based on the traditional spectroscopic method, i.e.,
excitation equilibrium of \ion{Fe}{1} and ionization equilibrium between \ion{Fe}{1} and
\ion{Fe}{2}. Their effective temperature is marginally consistent with that of
\citet{mishenina06} considering the uncertainty, but their surface gravity shows
relatively large discrepancy. Nevertheless, the adoption of stellar parameters for
HD\,83240 does not affect our validity test because difference in Si abundances determined
using the above two sets of stellar parameters is only 0.03\,dex.

\subsubsection{Non-LTE line formation results}

\begin{figure}
  \centering
  \includegraphics{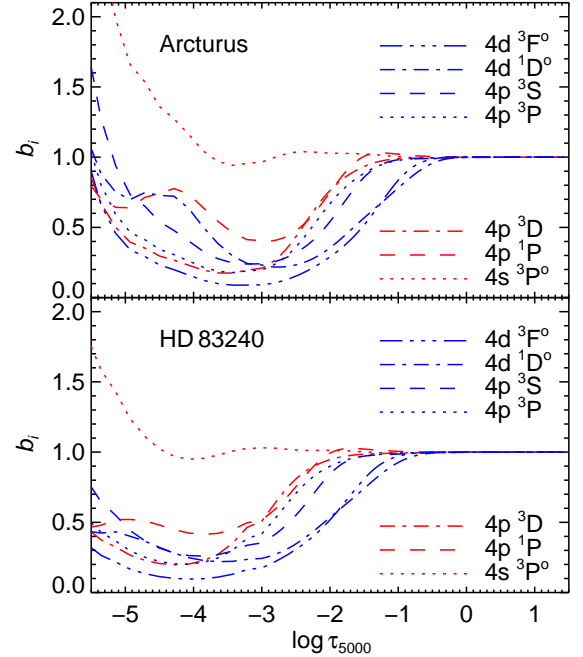}
  \caption{Departure coefficients for the populations of seven selected levels of \ion{Si}{1} as a function of continuum optical depth at
  5000\,{\AA} for Arcturus and HD\,83240. The red and the blue lines represent the lower and the upper levels, respectively, of the IR
  transitions used to determine Si abundances in this work.
  \label{fig:dep}}
\end{figure}

With the stellar parameters we could calculate the level populations in statistical
equilibrium. Figure~\ref{fig:dep} shows the departure coefficients
$b_i=n_i^{\mathrm{non\mhyphen LTE}}/n_i^{\mathrm{LTE}}$ for the populations of the seven
\ion{Si}{1} levels that produce the IR transitions used to determine Si abundances in this
work for the two benchmark stars. The red lines represent the lower levels, while the blue
lines are the upper levels. It can be seen that, within $\log\tau_{5000}=-3\ldots{-2}$,
where the \ion{Si}{1} IR lines are formed, the departure coefficients of the lower levels
(denoted with $b_i$) are higher than those of the upper levels (denoted with $b_j$). This
means that the line source function $S_{ij}$ is smaller than the local Planck function
$B_{\nu}(T)$ because $S_{ij}/B_{\nu}(T)\sim b_j/b_i<1$. As a result, the \ion{Si}{1} IR
lines are stronger in non-LTE than in LTE, and hence the non-LTE abundance corrections
should be negative.

\begin{figure}
  \centering
  \includegraphics{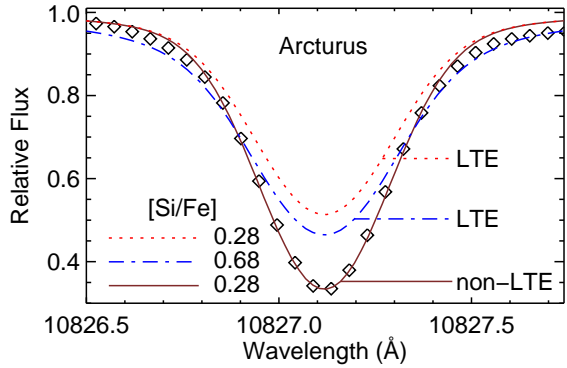}
  \caption{Spectrum synthesis of the \ion{Si}{1} 10827\,{\AA} line for Arcturus. The diamonds are the observed spectra; the lines are the
  synthetic spectra in LTE or non-LTE with different [Si/Fe] (see the legend for details).
  \label{fig:arcturus_syn}}
\end{figure}

In combination with the calculated populations for the individual levels, Si abundances
were then derived by spectrum synthesis of the individual \ion{Si}{1} lines. We note here
that, due to the enhanced absorption in the line cores in non-LTE, some of the \ion{Si}{1}
IR lines are so strong ($\mathrm{EW}\gtrsim150$\,m{\AA}) that their observed line profiles
could only be reproduced in non-LTE. As an example, Figure~\ref{fig:arcturus_syn} shows
the spectrum synthesis of the strongest \ion{Si}{1} IR line (10827\,{\AA} with a EW of
318.3\,m{\AA}) in Arcturus. The solid line shows the best fit to the observed line profile
in non-LTE with a [Si/Fe] of 0.28\,dex. The dotted line is produced with the same [Si/Fe]
but in LTE, which is much weaker in the core of the line. No matter how we adjusted the Si
abundance, the observed line profile cannot be reproduced in LTE. In this case, the LTE
abundance was derived by simply increasing the [Si/Fe] until the observed EW was
reproduced. This led to a [Si/Fe] of 0.68\,dex, for which the line profile is shown in
dash-dotted line in Figure~\ref{fig:arcturus_syn}. It can be seen that the core of the
line is still much shallower, while the wing is obviously deeper compared to the observed
spectra.

\floattable
\begin{deluxetable}{rrcccrrr}
\tablewidth{0pt}
\tablecolumns{8}
\tablecaption{EWs (in m{\AA}) and [Si/Fe] for the individual \ion{Si}{1} lines of the benchmark stars.\label{table:abun_bench}}
\tablehead{\colhead{Line} & \multicolumn{3}{c}{Arcturus} & & \multicolumn{3}{c}{HD\,83240}\\
\cline{2-4} \cline{6-8}
\colhead{({\AA})} & \colhead{EW} & \colhead{LTE} & \colhead{non-LTE} & & \colhead{EW} & \colhead{LTE} & \colhead{non-LTE}}
\startdata
 5690.43 &  59.1 & 0.29 & 0.29 & &  63.3 & $-0.09$ & $-0.09$\\
 5701.11 &  48.5 & 0.31 & 0.31 & &  54.1 & $-0.04$ & $-0.04$\\
 6142.49 &  31.2 & 0.29 & 0.29 & &  41.4 & $-0.02$ & $-0.02$\\
 6145.05 &  33.6 & 0.28 & 0.28 & &  44.6 & $-0.02$ & $-0.02$\\
  mean   &       & 0.29 & 0.29 & &       & $-0.04$ & $-0.04$\\
$\sigma$ &       & 0.01 & 0.01 & &       &   0.03  &   0.03 \\
\tableline
10288.90 &  82.3 & 0.40 & 0.25 & &  91.8 & $-0.01$ & $-0.07$\\
10371.30 & 160.9 & 0.66 & 0.26 & & 169.1 &   0.23  & $-0.06$\\
10585.17 & 257.1 & 0.76 & 0.30 & & 279.5 &   0.31  & $-0.01$\\
10603.45 & 216.8 & 0.68 & 0.23 & & 233.0 &   0.24  & $-0.06$\\
10627.66 & 102.4 & 0.59 & 0.18 & & 121.0 &   0.24  & $-0.11$\\
10661.00 & 226.8 & 0.74 & 0.28 & & 241.2 &   0.27  & $-0.02$\\
10689.73 & 126.5 & 0.66 & 0.30 & &\nodata& \nodata & \nodata\\
10694.27 & 135.0 & 0.66 & 0.26 & & 162.6 &   0.26  & $-0.02$\\
10727.43 & 153.3 & 0.68 & 0.25 & & 179.4 &   0.26  & $-0.06$\\
10749.40 & 237.0 & 0.68 & 0.22 & & 263.1 &   0.29  & $-0.03$\\
10784.57 &  68.4 & 0.44 & 0.29 & &\nodata& \nodata & \nodata\\
10786.88 & 223.0 & 0.70 & 0.24 & & 236.6 &   0.26  & $-0.06$\\
10827.10 & 318.3 & 0.68 & 0.28 & & 364.4 &   0.32  & $-0.01$\\
10843.87 & 141.6 & 0.72 & 0.24 & & 164.2 &   0.36  & $-0.07$\\
10882.83 &  64.0 & 0.36 & 0.22 & &  82.6 &   0.01  & $-0.08$\\
10979.34 & 197.7 & 0.68 & 0.25 & & 207.5 &   0.24  & $-0.06$\\
  mean   &       & 0.63 & 0.25 & &       &   0.23  & $-0.05$\\
$\sigma$ &       & 0.12 & 0.03 & &       &   0.11  &   0.03\\
\enddata
\end{deluxetable}

\begin{figure}
  \centering
  \includegraphics{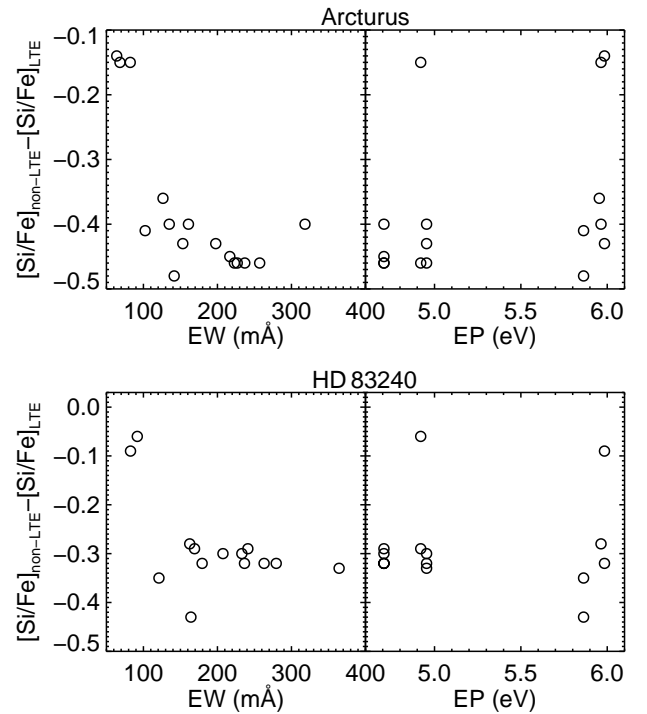}
  \caption{The non-LTE correction of [Si/Fe] for the individual \ion{Si}{1} IR lines as a function of EW and EP for Arcturus and HD\,83240.
  \label{fig:benchmark_nlte}}
\end{figure}

\begin{figure}
  \centering
  \includegraphics{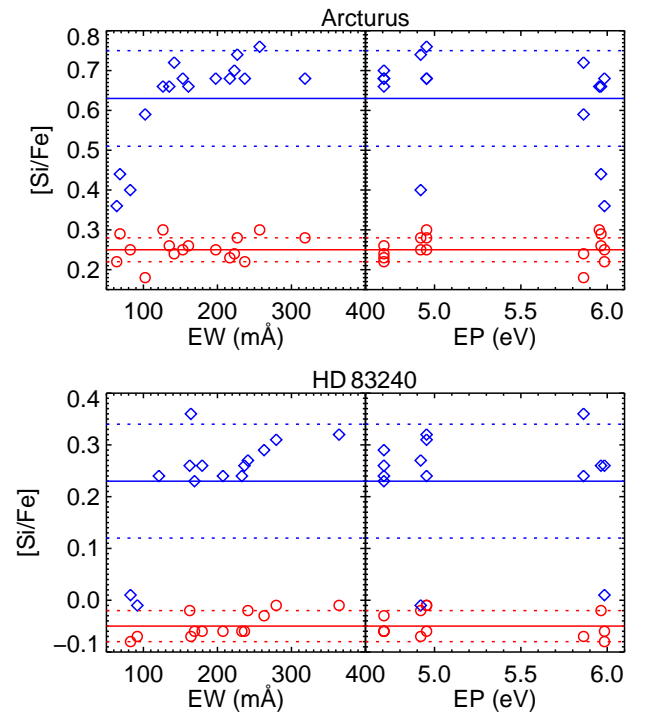}
  \caption{[Si/Fe] from the individual \ion{Si}{1} IR lines as a function of EW and EP for Arcturus and HD\,83240.
  The blue and the red are the LTE and the non-LTE results, respectively. The solid lines represent the average values, and the dotted lines
  indicate the $1\sigma$ lower and upper limits for the line-to-line scatter.
  \label{fig:benchmark_sife}}
\end{figure}

Table~\ref{table:abun_bench} gives [Si/Fe] determined from the individual
optical and IR lines for the two benchmark stars. EWs for the individual lines are also
given. As can be seen in Table~\ref{table:abun_bench}, for both Arcturus and HD\,83240,
the optical lines are insensitive to the non-LTE effects. This is because these lines are
mainly formed in the inner regions of the photospheres ($\log\tau_{5000}>-2$), where the
physical conditions are close to LTE. However, as we mentioned above, the regions of line
formation for the IR lines are shifted outward to $\log\tau_{5000}=-3\ldots{-2}$, where
the line source function differs from the local Planck function, and thus the non-LTE
effects cannot be neglected. Figure~\ref{fig:benchmark_nlte} shows the non-LTE correction
of [Si/Fe] for the individual \ion{Si}{1} IR lines as a function of EW and EP for the two
benchmark stars. It can be seen that the non-LTE abundance correction varies between
$\sim-0.5$\,dex and $\sim-0.05$\,dex. Though the non-LTE abundance correction seems to be
independent of EP, it is correlated with the EW of the line. In general, stronger lines
show larger non-LTE effects, while weaker lines show smaller non-LTE effects.
Figure~\ref{fig:benchmark_sife} shows the [Si/Fe] derived from the individual \ion{Si}{1}
IR lines as a function of EW and EP for the two benchmark stars. It is obvious that for
both Arcturus and HD\,83240, the LTE Si abundances from the IR lines show large line-to-line
scatter, but when the non-LTE effects are taken into account, the scatter reduces
significantly. Moreover, the average non-LTE Si abundances from the IR lines agree well
with those from the optical lines, whereas the LTE results are significantly higher. In a
word, our method of non-LTE calculations produces consistent Si abundances from the
optical and the IR lines for the two benchmark stars.

We were aware that the two benchmark stars we selected are relatively metal-rich,
and it is difficult to justify that our method works also for very metal-poor stars.
Unfortunately, in the publicly available archival database, we were not able to find any
very metal-poor giant star with high-quality IR spectra in the wavelength range
studied in this work. However, in a previous study, \citet{shi12} investigated Si abundances
for a well-studied very metal-poor giant HD\,122563 based on the IR spectra of
medium quality ($R\sim20,000$, $\mathrm{S/N}\sim100$). The stellar parameters they adopted
($T_{\mathrm{eff}}=4600$\,K, $\log g=1.5$, $\mathrm{[Fe/H]}=-2.53$) are in reasonable
agreement with the fundamental determinations ($T_{\mathrm{eff}}=4587\pm60$\,K,
$\log g=1.61\pm0.07$, $\mathrm{[Fe/H]}=-2.64\pm0.22$) of \citet{heiter15}. The Si abundance
determined from the IR spectra ([Si/Fe]$_{\mathrm{non\mhyphen LTE}}=0.20\pm0.01$)
agrees well with that from the 3905 and 4102\,{\AA} lines ([Si/Fe]$_{\mathrm{non\mhyphen LTE}}=0.22\pm0.05$).
Moreover, \citet{jofre15} determined Si abundance for HD\,122563 using five optical
lines\footnote{These lines are very weak; their EWs varies between 1 and 7\,m{\AA} according
to \citet{jofre15}.} by spectrum synthesis. They adopted the stellar parameters from
\citet{heiter15}, and their Si abundance ([Si/Fe]$_{\mathrm{non\mhyphen LTE}}=0.28\pm0.09$)
is consistent with those from \citet{shi12}.
Since we used exactly the same method as \citet{shi12}, it is fair to say that our method
works also for very metal-poor giant stars.

\section{Application to metal-poor giant stars}\label{appl}

In this section, we applied the above method of non-LTE calculations to a sample of 16
metal-poor giant stars.

\subsection{Stellar parameters and uncertainties}

The narrow wavelength range of our IR spectra prevented us from determining stellar
parameters for the sample stars by ourselves. So we adopted the stellar parameters from
\citet{takeda11,takeda12}, which were taken from nine published studies\footnote{It is
impossible to find independent measurements of stellar parameters for all the sample stars
in any single study.} (see Table~\ref{table:sample} for details). Among these studies,
different methods have been employed to determine stellar parameters. For example, some
studies derive $T_{\mathrm{eff}}$ based on spectroscopic method
(\citealt{fulbright00,simmerer04,hansen11}), while the others are based on photometric
calibrations; some studies determine $\log g$ using the parallaxes
(\citealt{simmerer04,takada05,saito09}), while the others utilize the ionization
equilibrium between \ion{Fe}{1} and \ion{Fe}{2}. Therefore, it is not surprising that
differences in stellar parameters exist between different studies. Even with the same
method, the adoption of different calibrations or line lists by different studies may lead
to different results (see the comparison of stellar parameters from different methods for
the FGK stars in the \emph{Gaia}-ESO survey by \citealt{smiljanic14}). In the nine studies
where we took the stellar parameters, 12 out of 16 sample stars have multiple independent
measurements of stellar parameters. We found that, for some stars, the uncertainties of stellar
parameters given by the original studies cannot interpret the large differences between
different studies. Therefore, for these 12 stars, we took the standard deviations of
stellar parameters between different studies as the uncertainties, which are given
in columns 7--10 of Table~\ref{table:sample}. For the rest four stars
(BD\,$+23\degr 3130$, HD\,13979, HD\,195636, and HE\,1523$-$0901) with only single
measurement of stellar parameters, the median value of the errors of the above 12 stars
was adopted.

\subsection{Si abundances and uncertainties}

\begin{figure}
  \centering
  \includegraphics{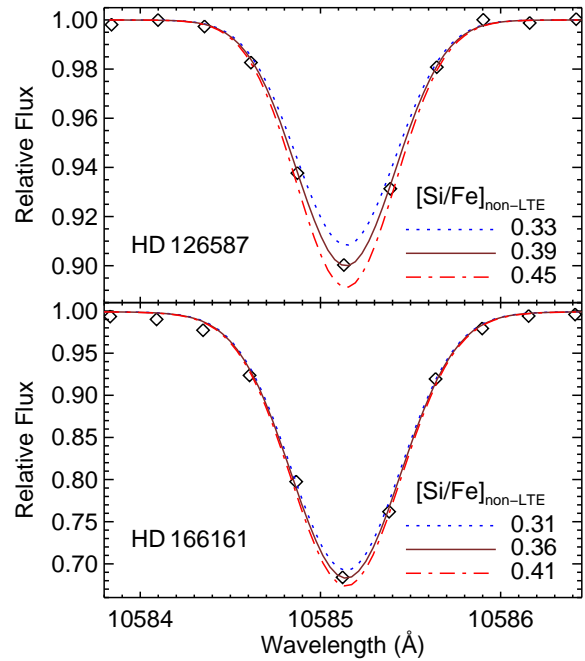}
  \caption{Spectrum synthesis of the \ion{Si}{1} 10585\,{\AA} line for HD\,126587 and HD\,166161. The diamonds are the observed spectra; the
  lines are the synthetic spectra in non-LTE with different [Si/Fe] (see the legend for details).
  \label{fig:si_syn}}
\end{figure}

Si abundances of the sample stars were derived by spectrum synthesis of the individual
\ion{Si}{1} IR lines. Figure~\ref{fig:si_syn} shows the spectrum synthesis of the strong
\ion{Si}{1} 10585\,{\AA} line for two program stars HD\,126587 and HD\,166161. HD\,126587
is the most metal-poor star in our sample, while HD\,166161 has the highest effective
temperature and surface gravity. It can be seen that, for both of these two stars, the
\ion{Si}{1} 10585\,{\AA} lines are strong enough ($\mathrm{EW}>60$\,m{\AA}) to be used to
determine Si abundances. Table~\ref{table:abun_line} gives [Si/Fe] determined from the
individual \ion{Si}{1} IR lines for the sample stars. EW for each line is also given.

\floattable
\begin{deluxetable}{lrrrrrrrrrrrrrr}
\tablewidth{0pt}
\tablecolumns{15}
\tablecaption{EWs (in m{\AA}) and [Si/Fe] for the individual \ion{Si}{1} IR lines of the 16 metal-poor giant stars.
For each star, the first row are the EWs; the second and the third row are the LTE and the non-LTE [Si/Fe], respectively.\label{table:abun_line}}
\tablehead{\colhead{Star} & \colhead{10371} & \colhead{10585} & \colhead{10603} & \colhead{10627} & \colhead{10661} & \colhead{10689} &
\colhead{10694} & \colhead{10727} &\colhead{10749} & \colhead{10784} & \colhead{10786} & \colhead{10827} & \colhead{10843} & \colhead{10979}}
\startdata
BD\,$+23\degr 3130$ &  23.1 &  77.1 &  54.2 &\nodata&  54.0 &\nodata&\nodata&\nodata&  65.7 &\nodata&  55.0 & 114.4 &\nodata&  39.7 \\
                    &  0.38 &  0.58 &  0.40 &\nodata&  0.38 &\nodata&\nodata&\nodata&  0.41 &\nodata&  0.40 &  0.69 &\nodata&  0.37 \\
                    &  0.16 &  0.16 &  0.14 &\nodata&  0.12 &\nodata&\nodata&\nodata&  0.11 &\nodata&  0.13 &  0.14 &\nodata&  0.16 \\
BD\,$-16\degr 251$  &  27.5 &  84.5 &  57.2 &\nodata&  60.9 &\nodata&\nodata&\nodata&  66.7 &\nodata&  62.0 & 126.7 &\nodata&\nodata\\
                    &  0.68 &  0.83 &  0.63 &\nodata&  0.66 &\nodata&\nodata&\nodata&  0.60 &\nodata&  0.67 &  1.00 &\nodata&\nodata\\
                    &  0.39 &  0.35 &  0.27 &\nodata&  0.29 &\nodata&\nodata&\nodata&  0.21 &\nodata&  0.30 &  0.29 &\nodata&\nodata\\
BD\,$-18\degr 5550$ &  24.3 &  79.4 &  47.9 &\nodata&\nodata&\nodata&\nodata&\nodata&  70.6 &\nodata&  49.5 & 110.9 &\nodata&\nodata\\
                    &  0.72 &  0.87 &  0.60 &\nodata&\nodata&\nodata&\nodata&\nodata&  0.76 &\nodata&  0.61 &  0.88 &\nodata&\nodata\\
                    &  0.43 &  0.41 &  0.26 &\nodata&\nodata&\nodata&\nodata&\nodata&  0.34 &\nodata&  0.26 &  0.26 &\nodata&\nodata\\
HD\,6268            &  59.7 & 125.4 &  86.4 &  13.8 &  98.4 &  26.3 &  35.1 &  41.4 & 107.0 &\nodata&  89.9 & 158.8 &\nodata&\nodata\\
                    &  0.50 &  0.61 &  0.30 &  0.26 &  0.42 &  0.37 &  0.41 &  0.34 &  0.39 &\nodata&  0.31 &  0.57 &\nodata&\nodata\\
                    &  0.27 &  0.11 &  0.03 &  0.21 &  0.11 &  0.21 &  0.24 &  0.16 &  0.04 &\nodata&  0.04 &  0.00 &\nodata&\nodata\\
HD\,13979           &\nodata&  92.3 &  57.0 &\nodata&  64.1 &\nodata&  23.3 &  29.5 &  73.2 &\nodata&  58.9 & 120.8 &\nodata&  43.3 \\
                    &\nodata&  0.57 &  0.18 &\nodata&  0.27 &\nodata&  0.28 &  0.25 &  0.28 &\nodata&  0.19 &  0.60 &\nodata&  0.14 \\
                    &\nodata&  0.06 &$-0.07$&\nodata&$-0.01$&\nodata&  0.09 &  0.04 &$-0.05$&\nodata&$-0.07$&$-0.02$&\nodata&$-0.04$\\
HD\,108317          &  33.7 &  99.4 &  65.7 &\nodata&  71.7 &\nodata&\nodata&  41.2 &  87.1 &\nodata&  72.9 & 125.9 &\nodata&  44.4 \\
                    &  0.49 &  0.69 &  0.43 &\nodata&  0.49 &\nodata&\nodata&  0.58 &  0.53 &\nodata&  0.49 &  0.60 &\nodata&  0.33 \\
                    &  0.29 &  0.26 &  0.19 &\nodata&  0.23 &\nodata&\nodata&  0.39 &  0.23 &\nodata&  0.24 &  0.12 &\nodata&  0.16 \\
HD\,115444          &  21.7 &  76.6 &  46.9 &\nodata&  54.8 &\nodata&\nodata&\nodata&  58.3 &\nodata&  58.7 & 100.7 &\nodata&\nodata\\
                    &  0.29 &  0.43 &  0.21 &\nodata&  0.30 &\nodata&\nodata&\nodata&  0.16 &\nodata&  0.35 &  0.31 &\nodata&\nodata\\
                    &  0.09 &  0.08 &$-0.01$&\nodata&  0.06 &\nodata&\nodata&\nodata&$-0.04$&\nodata&  0.10 &$-0.09$&\nodata&\nodata\\
HD\,121135          & 115.6 & 218.9 & 172.4 &  70.2 & 186.1 &  97.4 & 114.1 & 129.6 & 194.9 &  34.9 & 185.8 &\nodata& 118.0 &\nodata\\
                    &  0.55 &  1.06 &  0.76 &  0.48 &  0.88 &  0.62 &  0.71 &  0.74 &  0.83 &  0.32 &  0.88 &\nodata&  0.79 &\nodata\\
                    &  0.21 &  0.36 &  0.21 &  0.23 &  0.30 &  0.33 &  0.36 &  0.32 &  0.23 &  0.24 &  0.29 &\nodata&  0.36 &\nodata\\
HD\,126587          &\nodata&  69.1 &  48.7 &\nodata&  52.0 &\nodata&\nodata&\nodata&  53.4 &\nodata&  48.5 & 114.3 &\nodata&\nodata\\
                    &\nodata&  0.82 &  0.70 &\nodata&  0.73 &\nodata&\nodata&\nodata&  0.61 &\nodata&  0.68 &  1.05 &\nodata&\nodata\\
                    &\nodata&  0.39 &  0.33 &\nodata&  0.35 &\nodata&\nodata&\nodata&  0.23 &\nodata&  0.31 &  0.36 &\nodata&\nodata\\
HD\,166161          & 143.9 & 243.0 & 204.1 &  83.2 & 206.3 & 111.6 & 128.9 & 151.6 & 221.3 &  45.6 & 189.7 & 288.0 & 133.2 & 193.2 \\
                    &  0.76 &  1.01 &  0.86 &  0.54 &  0.86 &  0.61 &  0.68 &  0.74 &  0.84 &  0.44 &  0.68 &  0.88 &  0.74 &  0.91 \\
                    &  0.40 &  0.36 &  0.36 &  0.36 &  0.35 &  0.36 &  0.37 &  0.36 &  0.31 &  0.36 &  0.23 &  0.29 &  0.41 &  0.46 \\
HD\,186478          &  48.7 & 125.2 &  87.4 &\nodata&  96.6 &\nodata&\nodata&  45.4 & 103.4 &\nodata&  89.8 & 159.1 &\nodata&\nodata\\
                    &  0.50 &  0.82 &  0.49 &\nodata&  0.59 &\nodata&\nodata&  0.55 &  0.52 &\nodata&  0.49 &  0.85 &\nodata&\nodata\\
                    &  0.26 &  0.25 &  0.16 &\nodata&  0.22 &\nodata&\nodata&  0.33 &  0.14 &\nodata&  0.16 &  0.16 &\nodata&\nodata\\
HD\,195636          &\nodata&  65.4 &  40.9 &\nodata&  44.4 &\nodata&\nodata&\nodata&\nodata&\nodata&\nodata&\nodata&\nodata&\nodata\\
                    &\nodata&  0.75 &  0.54 &\nodata&  0.58 &\nodata&\nodata&\nodata&\nodata&\nodata&\nodata&\nodata&\nodata&\nodata\\
                    &\nodata&  0.32 &  0.23 &\nodata&  0.26 &\nodata&\nodata&\nodata&\nodata&\nodata&\nodata&\nodata&\nodata&\nodata\\
HD\,204543          & 100.3 & 216.3 & 165.0 &  45.5 & 181.1 &  66.5 &  73.4 & 104.0 & 173.0 &\nodata& 157.9 & 237.7 &  89.2 &\nodata\\
                    &  0.43 &  1.09 &  0.69 &  0.34 &  0.86 &  0.41 &  0.38 &  0.58 &  0.62 &\nodata&  0.56 &  0.84 &  0.58 &\nodata\\
                    &  0.15 &  0.32 &  0.16 &  0.16 &  0.26 &  0.21 &  0.16 &  0.26 &  0.06 &\nodata&  0.08 &  0.10 &  0.30 &\nodata\\
HD\,216143          &  64.4 & 145.5 & 111.6 &\nodata& 121.0 &\nodata&  47.6 &  61.2 & 131.9 &\nodata& 115.4 & 183.5 &  40.2 &\nodata\\
                    &  0.17 &  0.46 &  0.21 &\nodata&  0.31 &\nodata&  0.25 &  0.27 &  0.31 &\nodata&  0.24 &  0.47 &  0.14 &\nodata\\
                    &  0.01 &  0.00 &$-0.05$&\nodata&  0.01 &\nodata&  0.10 &  0.09 &$-0.04$&\nodata&$-0.04$&$-0.07$&  0.00 &\nodata\\
HD\,221170          &  83.4 & 168.9 & 126.8 &  31.7 & 134.9 &  48.6 &  57.5 &  74.1 & 155.0 &\nodata& 129.8 & 201.1 &  57.1 &\nodata\\
                    &  0.50 &  0.94 &  0.58 &  0.37 &  0.66 &  0.44 &  0.46 &  0.51 &  0.78 &\nodata&  0.58 &  0.86 &  0.45 &\nodata\\
                    &  0.23 &  0.26 &  0.13 &  0.22 &  0.18 &  0.26 &  0.25 &  0.26 &  0.21 &\nodata&  0.13 &  0.16 &  0.26 &\nodata\\
HE\,1523$-$0901     &  23.6 &  83.1 &  62.9 &\nodata&  73.1 &\nodata&\nodata&\nodata&  81.4 &\nodata&  67.2 & 127.3 &\nodata&\nodata\\
                    &  0.51 &  0.62 &  0.56 &\nodata&  0.65 &\nodata&\nodata&\nodata&  0.61 &\nodata&  0.59 &  0.68 &\nodata&\nodata\\
                    &  0.26 &  0.24 &  0.24 &\nodata&  0.31 &\nodata&\nodata&\nodata&  0.24 &\nodata&  0.26 &  0.14 &\nodata&\nodata\\
\enddata
\end{deluxetable}

We were aware that stellar parameters for the sample stars were adopted from different
studies, and the effect of inhomogeneous stellar parameters on Si abundances needs to be
properly estimated. To do this, for the 12 stars with multiple determinations of stellar
parameters, we calculated Si abundances from different stellar parameters, and took the
standard deviations as the uncertainties of Si abundances caused by stellar parameters.
For the rest four stars with only single measurement of stellar parameters, the median
value of the errors (0.12\,dex) of the above 12 stars was adopted as the uncertainties
caused by stellar parameters. The total errors of [Si/Fe] were then calculated by adding
the line-to-line scatters and the errors introduced by stellar parameters in quadrature.
Table~\ref{table:abun} gives the average [Si/Fe] from the \ion{Si}{1} IR lines as well as
the associated uncertainties for all the stars investigated in this work.

\floattable
\begin{deluxetable}{lcrcccccc}
\tablewidth{0pt}
\tablecolumns{9}
\tablecaption{Si abundances and uncertainties for all the sample stars. The total errors ($\sigma_{\mathrm{total}}$) were calculated
by adding the line-to-line scatters ($\sigma_{\mathrm{line}}$) and the errors introduced by stellar parameters ($\sigma_{\mathrm{par}}$)
in quadrature.\label{table:abun}}
\tablehead{\colhead{Star} & \multicolumn{2}{c}{[Si/Fe]} &  & \multicolumn{2}{c}{$\sigma_{\mathrm{line}}$} & \colhead{$\sigma_{\mathrm{par}}$} & \multicolumn{2}{c}{$\sigma_{\mathrm{total}}$}\\
\cline{2-3} \cline{5-6} \cline{8-9}
 & \colhead{LTE} & \colhead{non-LTE} &  & \colhead{LTE} & \colhead{non-LTE} &  & \colhead{LTE} & \colhead{non-LTE}}
\startdata
Arcturus            & 0.63 &   0.25  & & 0.12 & 0.03 & 0.03 & 0.12 & 0.04\\
HD\,83240           & 0.23 & $-0.05$ & & 0.11 & 0.03 & 0.02 & 0.11 & 0.04\\
\tableline
BD\,$+23\degr 3130$ & 0.45 &   0.14  & & 0.12 & 0.02 & 0.12 & 0.17 & 0.12\\
BD\,$-16\degr 251$  & 0.72 &   0.30  & & 0.14 & 0.06 & 0.01 & 0.14 & 0.06\\
BD\,$-18\degr 5550$ & 0.74 &   0.33  & & 0.12 & 0.08 & 0.06 & 0.13 & 0.10\\
HD\,6268            & 0.41 &   0.13  & & 0.11 & 0.09 & 0.44 & 0.45 & 0.45\\
HD\,13979           & 0.31 & $-0.01$ & & 0.17 & 0.06 & 0.12 & 0.21 & 0.13\\
HD\,108317          & 0.51 &   0.23  & & 0.10 & 0.08 & 0.07 & 0.12 & 0.11\\
HD\,115444          & 0.29 &   0.03  & & 0.09 & 0.07 & 0.12 & 0.15 & 0.14\\
HD\,121135          & 0.72 &   0.29  & & 0.20 & 0.06 & 0.05 & 0.21 & 0.08\\
HD\,126587          & 0.77 &   0.33  & & 0.16 & 0.06 & 0.12 & 0.20 & 0.13\\
HD\,166161          & 0.75 &   0.36  & & 0.15 & 0.05 & 0.14 & 0.21 & 0.15\\
HD\,186478          & 0.60 &   0.21  & & 0.15 & 0.07 & 0.16 & 0.22 & 0.17\\
HD\,195636          & 0.62 &   0.27  & & 0.11 & 0.05 & 0.12 & 0.16 & 0.13\\
HD\,204543          & 0.62 &   0.19  & & 0.22 & 0.09 & 0.07 & 0.23 & 0.11\\
HD\,216143          & 0.28 &   0.00  & & 0.11 & 0.06 & 0.11 & 0.16 & 0.13\\
HD\,221170          & 0.59 &   0.21  & & 0.18 & 0.05 & 0.16 & 0.24 & 0.17\\
HE\,1523$-$0901     & 0.60 &   0.24  & & 0.06 & 0.05 & 0.12 & 0.13 & 0.13\\
\enddata
\end{deluxetable}

\subsection{Non-LTE effects of the \ion{Si}{1} IR lines}

\begin{figure}
  \centering
  \includegraphics{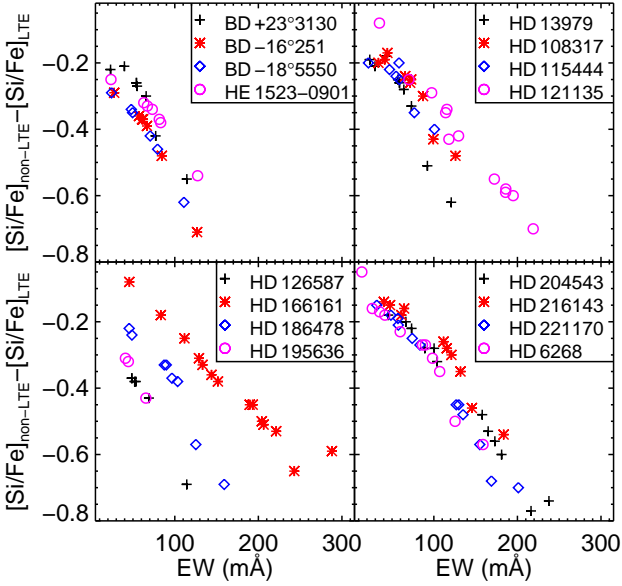}
  \caption{The non-LTE correction of [Si/Fe] for the individual \ion{Si}{1} IR lines as a function of EW for the 16 metal-poor giant stars.
  \label{fig:nlte_ew}}
\end{figure}

\begin{figure}
  \centering
  \includegraphics{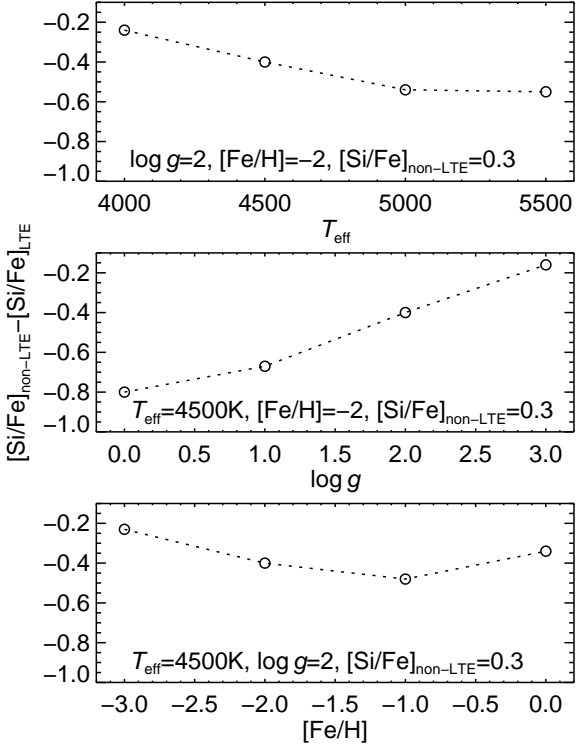}
  \caption{The non-LTE correction of [Si/Fe] for the \ion{Si}{1} 10603\,{\AA} line as a function of stellar parameters.
  \label{fig:nlte_par}}
\end{figure}

\begin{figure}
  \centering
  \includegraphics{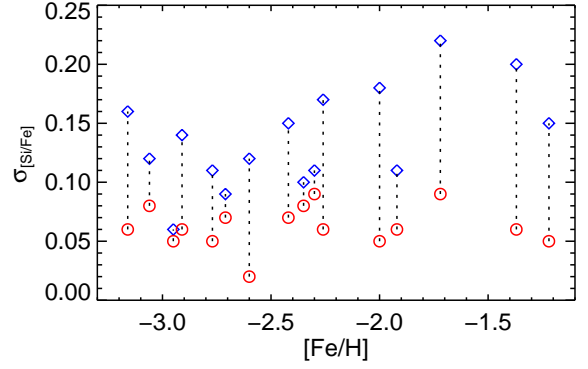}
  \caption{The line-to-line scatter of [Si/Fe] for the 16 metal-poor giant stars. The diamonds and the circles are the LTE and the non-LTE
  results, respectively.
  \label{fig:line_scatter}}
\end{figure}

\begin{figure}
  \centering
  \includegraphics{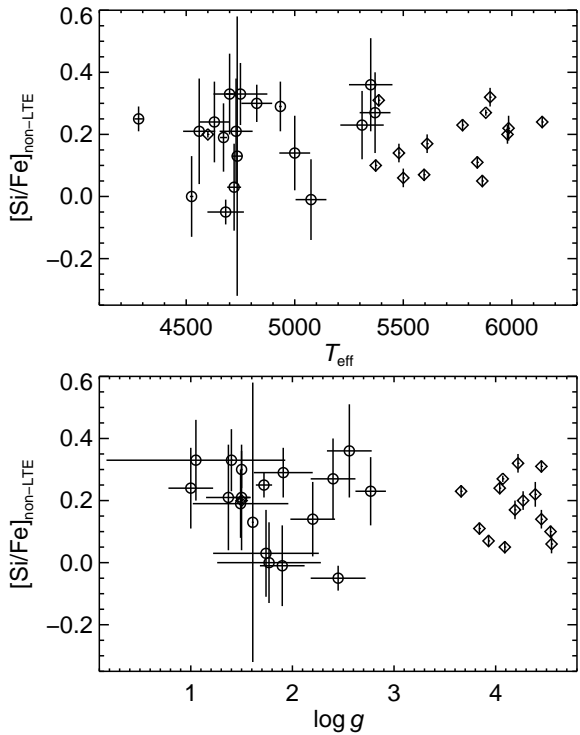}
  \caption{[Si/Fe] derived in non-LTE as a function of effective temperature and surface gravity. The circles are our results; the diamonds are
  the results from \citet{shi12}.
  \label{fig:si_par}}
\end{figure}

Figure~\ref{fig:nlte_ew} shows the non-LTE correction of [Si/Fe] for the individual
\ion{Si}{1} IR lines as a function of EW for the sample stars. It can be seen that, the
non-LTE effects are significant for all the stars. Besides, the non-LTE abundance
corrections differ from line to line; in general, stronger lines show larger non-LTE
effects (up to $\sim0.8$\,dex), while weaker lines show smaller non-LTE effects (as low as
$\sim0.1$\,dex). Overall, the non-LTE abundance corrections show linear correlations with
the line strengths for all the stars with similar slopes. But the intercepts of the linear
relationships are not necessarily the same; they are dependent on the stellar parameters,
especially the surface gravity (see the discussions below). We also explored the
dependency of the non-LTE effects on the stellar parameters, and the results are plotted
in Figure~\ref{fig:nlte_par}. It can be seen that, the non-LTE abundance corrections are
all negative within the range of stellar parameters investigated. Moreover, the non-LTE
abundance corrections are more sensitive to surface gravity than effective temperature and
metallicity. Stars with lower surface gravity show larger non-LTE effects, while stars
with higher gravity show smaller non-LTE effects. Figure~\ref{fig:line_scatter} shows the
line-to-line scatter of [Si/Fe] for the sample stars. It can be seen that, for a given
star, the LTE Si abundances show relatively large line-to-line scatter. However, when the
non-LTE effects are considered, the scatter reduces significantly, though for some stars
they are still relatively large compared to the benchmark stars. This should be partly due
to the quality of the spectra (lower resolution and lower S/N compared to the spectra of
the benchmark stars) used for abundance determination. Figure~\ref{fig:si_par} shows
[Si/Fe] as a function of effective temperature and surface gravity. The results from
\citet{shi12} which are based on the same IR lines and the same method are also plotted.
It can be seen that [Si/Fe] derived from the \ion{Si}{1} IR lines based on the non-LTE
analysis is independent of $T_{\mathrm{eff}}$ and $\log g$.

\subsection{Comparison with other studies}

\citet{shi12} investigated the non-LTE effects of the \ion{Si}{1} IR lines for 15 nearby
stars. Their sample are mostly dwarf stars, and there is only one giant star (HD\,122563),
for which the average non-LTE correction of [Si/Fe] is $-0.26$\,dex. This is consistent
with the results of our sample stars with similar stellar parameters.

\citet{bergemann13} calculated the non-LTE effects of four \ion{Si}{1} IR lines for RSG
stars with effective temperature between 3400 and 4400\,K. Unfortunately, we are not able
to compare our results directly with those of \citet{bergemann13} because the \ion{Si}{1}
IR lines they investigated are different from those this work. Nevertheless, their results
also show negative non-LTE abundance correction, though the magnitude of correction
($-0.4\ldots{-0.1}$\,dex) is lower than our results. This should be mainly due to that
\citet{bergemann13} adopted stronger inelastic collisions with neutral hydrogen, which
leads to smaller non-LTE effects.

\subsection{Implications for the chemical evolution of Si}

\begin{figure}
  \centering
  \includegraphics{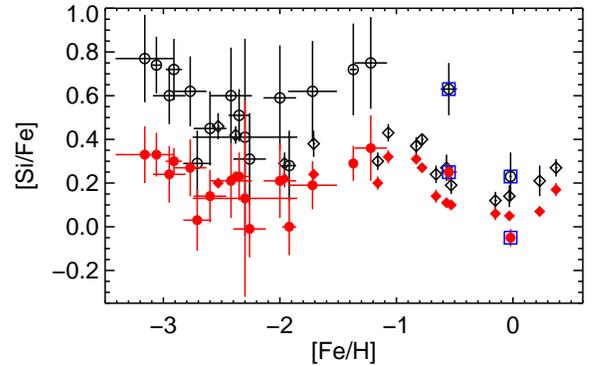}
  \caption{[Si/Fe] as a function of [Fe/H]. Our results are plotted as circles; the results from \citet{shi12} are plotted as diamonds. Open and
  filled symbols correspond to the LTE and the non-LTE abundances, respectively. The results for the two benchmark stars are marked with squares.
  \label{fig:sife}}
\end{figure}

As we mentioned in the introduction, Si abundances could be used to test the SNe and
Galactic chemical evolution models. Several previous studies on stellar abundances have
investigated the relationship between [Si/Fe] and [Fe/H]
(\citealt{fulbright00,cayrel04,shi09}; etc.). While it is generally accepted that [Si/Fe]
decreases with [Fe/H] from $\mathrm{[Fe/H]}\sim-1$ to $\mathrm{[Fe/H]}\sim0$, the behavior
of [Si/Fe] below $\mathrm{[Fe/H]}\sim-1$ is still controversial. One reason for this is
the significant scatter of [Si/Fe] (ranging from $-0.2$ to 0.9\,dex; see Figure~1 of
\citealt{shi09}) below $\mathrm{[Fe/H]}\sim-1$. The dispersion could be due to either the
cosmic scatter or the uncertainties in abundance determinations or both.
\citet{fulbright00} determined Si abundances in LTE for 168 stars ($-3<\mathrm{[Fe/H]}<0$)
using 12 \ion{Si}{1} lines between 5600 and 7100\,{\AA}. Their results showed a decreasing
trend of [Si/Fe] with increasing [Fe/H]. \citet{cayrel04} performed abundance analysis for
35 very metal-poor stars. Their Si abundances were derived in LTE using the \ion{Si}{1}
4102\,{\AA} line. The results indicated a slightly increasing trend of [Si/Fe] from
$\mathrm{[Fe/H]}\sim-4$ to $\mathrm{[Fe/H]}\sim-2$. \citet{shi09,shi11} determined Si
abundances in non-LTE for 79 stars based on 11 \ion{Si}{1} lines between 3900 and
6300\,{\AA} as well as two \ion{Si}{2} lines (6347/6371\,{\AA}). Their results suggested a
flat trend of [Si/Fe] from $\mathrm{[Fe/H]}\sim-3$ to $\mathrm{[Fe/H]}\sim-1$, and then a
decreasing trend above $\mathrm{[Fe/H]}\sim-1$. The trend between [Si/Fe] and [Fe/H]
derived in this work is shown in Figure~\ref{fig:sife}. The results from \citet{shi12}
which are obtained using the same IR lines and the same method are also plotted. All the
stars in our sample are giants, while most of the stars from \citet{shi12} are dwarfs
(only one giant). It can be seen that there is no difference in [Si/Fe] between giants and
dwarfs in the common metallicity range. As shown in Figure~\ref{fig:sife}, combination of
our results with those from \citet{shi12} suggests that [Si/Fe] decreases with increasing
[Fe/H] in general, though there seems to be a bump around $\mathrm{[Fe/H]}\sim-1$.

There are several Galactic chemical evolution models referring to Si in the literature
(\citealt{timmes95,samland98,kobayashi11}; etc.). The chemical evolution calculation of
\citet{timmes95} predicted that [Si/Fe] increases with [Fe/H] from $\mathrm{[Fe/H]}\sim-3$
to $\mathrm{[Fe/H]}\sim-2$, and then decreases with [Fe/H] until $\mathrm{[Fe/H]}\sim-0.8$.
\citet{samland98} predicted a constant ratio of [Si/Fe] in the early Galaxy and then a
decreasing trend with increasing [Fe/H] due to the contribution of SNe~Ia.
\citet{kobayashi11} presented the evolution of elements (from C to Zn) using chemical
evolution models with updated yields of asymptotic giant branch (AGB) stars and
core-collapse SNe. Their results showed that [Si/Fe] gradually decreases with [Fe/H] from
$\mathrm{[Fe/H]}\sim-4$ to $\mathrm{[Fe/H]}\sim-1$. Above $\mathrm{[Fe/H]}\sim-1$, [Si/Fe]
decreases more rapidly with increasing [Fe/H] due to the contribution of SNe~Ia.
Interestingly, if the effects of rotating massive stars at $Z=0$ (in addition to
hypernovae, SNe~II, SNe~Ia, and AGB stars) are taken into account, the [Si/Fe]
\emph{versus} [Fe/H] trend that they predicted also shows a small bump around
$\mathrm{[Fe/H]}\sim-1$. Such bumps are also predicted for other $\alpha$-elements, such
as S, Ca, and Ti. Therefore, our observational results are most consistent with the
predictions of \citet{kobayashi11}. However, due to the inhomogeneous stellar parameters
and small number of sample stars, we are not able to make any conclusive remarks on the
Galactic chemical evolution of Si.

\section{Summary}\label{sum}

As an important $\alpha$-element, Si is believed to be mainly produced by SNe~II, but it
is not clear whether SNe~Ia also produces some Si. Therefore, Si abundances could be used
to test the SNe and Galactic chemical evolution models. Unfortunately, the optical
\ion{Si}{1} lines are very weak in very metal-poor stars, and the NUV \ion{Si}{1} lines
are either blended or very difficult to be normalized. In this regard, the \ion{Si}{1}
IR lines could be better abundance indicators because they are usually much stronger than
the optical lines and they suffer much less from the problem of blending or continuum
normalization compared to the NUV lines. However, LTE is not a realistic approximation for
the line formation of the \ion{Si}{1} IR lines, so we have investigated the non-LTE
effects of the \ion{Si}{1} IR lines in giant stars. The main results could be summarized
as follows.
\begin{itemize}
  \item Si abundances based on the LTE analysis of the \ion{Si}{1} IR lines are
  overestimated (typical value of $\sim0.35$\,dex for giant stars), and thus are higher
  than those from the optical lines which are insensitive to the non-LTE effects.
  However, when our non-LTE calculations are applied, Si abundances from the optical and
  the IR lines are consistent.
  \item The non-LTE effects of the \ion{Si}{1} infrared lines differ from line to line.
  In general, stronger lines show larger non-LTE effects (up to $\sim0.8$\,dex) while
  weaker lines show smaller non-LTE effects (as low as $\sim0.1$\,dex). Therefore, it is
  not surprising that Si abundances based on the LTE analysis of the \ion{Si}{1} IR lines
  show large line-to-line scatter (mean value of 0.13\,dex), and when our non-LTE
  calculations are applied, the scatter reduces significantly (mean value of 0.06\,dex).
  \item The non-LTE effects of the \ion{Si}{1} infrared lines are dependent on stellar
  parameters, among which the surface gravity plays a dominant role. Giant stars show
  larger non-LTE effects (typical value of $\sim0.35$\,dex), while dwarf stars show
  smaller non-LTE effects (typical value of $\sim0.1$\,dex).
\end{itemize}
Therefore, the \ion{Si}{1} IR lines could be reliable abundance indicators provided
that the non-LTE effects are properly taken into account. Our results remind that one
should be very careful when using the IR lines to determine chemical abundances under the
assumption of LTE. In particular, the APOGEE/APOGEE-2 project will provide high-resolution
and high S/N ratio spectra in the $H$-band for about 400,000 stars, based on which
abundances of up to 15 chemical species could be obtained. Investigations of the non-LTE
effects for the $H$-band spectra lines for these elements is of great importance to
improving the accuracy of abundance determinations.


\acknowledgments

We are grateful to the anonymous referee for the valuable suggestions and comments that
improve the paper substantially.
This work is supported by the National Nature Science Foundation of China under grant
Nos.~11103034, 11321064, 11233004, 11390371, 11473033, 11428308, 11273002, U1331120,
U1331122, and by the National Basic Research Program of China under grant No.~2014CB845701.
This research has made use of the SIMBAD database, operated at CDS, Strasbourg, France,
and the NASA's Astrophysics Data System.

\end{document}